\documentclass[floatfix,superscriptaddress]{revtex4-2}

\usepackage{graphicx}
\usepackage{dcolumn}
\usepackage{bm}

\begin{document}

\preprint{APS/123-QED}

\title{Hydride superconductivity: here to stay}

\author{Gregory S. Boebinger} 
\affiliation{Florida State University, Tallahassee, FL 32306 USA}
\author{Andrey V. Chubukov}
\affiliation{University of Minnesota, School of Physics and  Astronomy, Minneapolis, MN 55455 USA}
\author{Ian R. Fisher}
\affiliation{Geballe Laboratory for Advanced Materials and Department of Applied Physics, Stanford University, CA, USA}
\affiliation{Stanford Institute for Materials and Energy Sciences, SLAC, Menlo Park, CA, USA}
\author{F. Malte Grosche}
\affiliation{University of Cambridge, Cavendish Laboratory, Cambridge CB3 0HE, England}
\author{Peter J. Hirschfeld}
\affiliation{University of Florida, Dept Phys, Gainesville, FL 32603 USA}
\author{Stephen R. Julian}
\affiliation{University of Toronto, Dept Phys, Toronto, ON M5S 1A7, Canada}
\author{Bernhard Keimer}
\affiliation{Max Planck Institute for Solid State Research, Stuttgart, Germany}
\author{Steven A. Kivelson}
\affiliation{Stanford University, Dept Phys, Stanford, CA 94305 USA}
\author{Andrew P. Mackenzie}
\affiliation{Max Planck Institute for Chemical Physics of Solids, Nothnitzer Str 40, D-01187 Dresden, Germany}
\affiliation{University St Andrews, Sch Phys \& Astron, SUPA, St Andrews KY16 9SS, Fife, Scotland}
\author{Yoshiteru Maeno}
\affiliation{Kyoto University, Toyota Riken Kyoto Univ Res Ctr, Kyoto 6068501, Japan}
\author{Joseph Orenstein}
\affiliation{University of California at Berkeley, Dept Phys, Berkeley, CA 94720 USA}
\affiliation{Lawrence Berkeley National Laboratory, Mat Sci Div, Berkeley, CA 94720 USA}
\author{Brad J. Ramshaw}
\affiliation{Cornell University, Lab Atom \& Solid State Phys, Ithaca, NY 14850 USA}
\author{Subir Sachdev}
\affiliation{Harvard University, Dept Phys, Cambridge, MA 02138 USA}
\author{J\"org Schmalian}
\affiliation{Karlsruhe Institute of Technology, D-76131 Karlsruhe, Germany}
\author{Matthias Vojta}
\affiliation{Technical University of Dresden, D-01062 Dresden, Germany}

\date{\today}

\begin{abstract}
The field of hydride superconductivity has recently been mired in a controversy that might divert attention from the question of central importance: do hydrides support genuine superconductivity or not? We examine some key papers from the field, and conclude that hydride superconductivity is real.
\end{abstract}

\maketitle


\section*{Introduction}
Superconductivity is one of nature's most captivating phenomena. In this state, a superconductor achieves macroscopic quantum coherence, allowing it to carry electrical currents indefinitely without energy loss.  Investigating new potential superconductors is often highly challenging, especially during the initial months and years after discovery, when material purity is typically low. Even now well-established superconductors initially exhibited ambiguous signs of this state. On the other hand, some widely heralded high-temperature superconductors were - after scrutiny by the research community - shown to have been misidentified.

One of the forefront fields of modern superconductivity research is that on hydrides at high pressures.  The past few years have seen this research attract considerable publicity, of which a substantial fraction has been negative.  Scientific fraud has been committed and exposed, and arguments continue about specific aspects of data presented in some other papers.  Among all the noise that is being generated, one might lose sight of the big-picture question of whether the field is on solid foundations or not, i.e. whether high-pressure hydrides host superconductivity at all.  Here, we readdress this central issue.  We select and critically examine what we identify as six key papers on the topic.  We have all spent substantial portions of our careers working on superconductivity, so hope that the conclusions that we reach will carry at least some weight with interested but non-specialist readers.  We also decided to include among our authorship team only people who have never worked directly on hydride superconductivity, so that our examination of the scientific facts can be as impartial as possible. We conclude that it is overwhelmingly probable, based on the body of scientific evidence that has been accumulated, that the phenomenon of hydride superconductivity is genuine. 

\section*{Challenges of hydride sample preparation and physical property measurement}

What most people would regard as the breakthrough paper in hydride superconductivity was published in 2015 by the group at the Max Planck Institute in Mainz\cite{1}.  It described the extreme compression of H$_2$S to pressures over one million times higher than that of our atmosphere, and the observation of a transition reported to be superconductivity at approximately $200\, {\rm K}$, attributed to the formation of H$_3$S.  This famous work was subsequently followed by other experiments in which mixtures of hydrogen and metal atoms were both compressed and heated, creating in situ chemical reactions to form compounds directly in the pressure cells, which were subsequently cooled to check for signs of superconductivity.  
\begin{figure}[ht]
\centering
\includegraphics[scale=0.50]{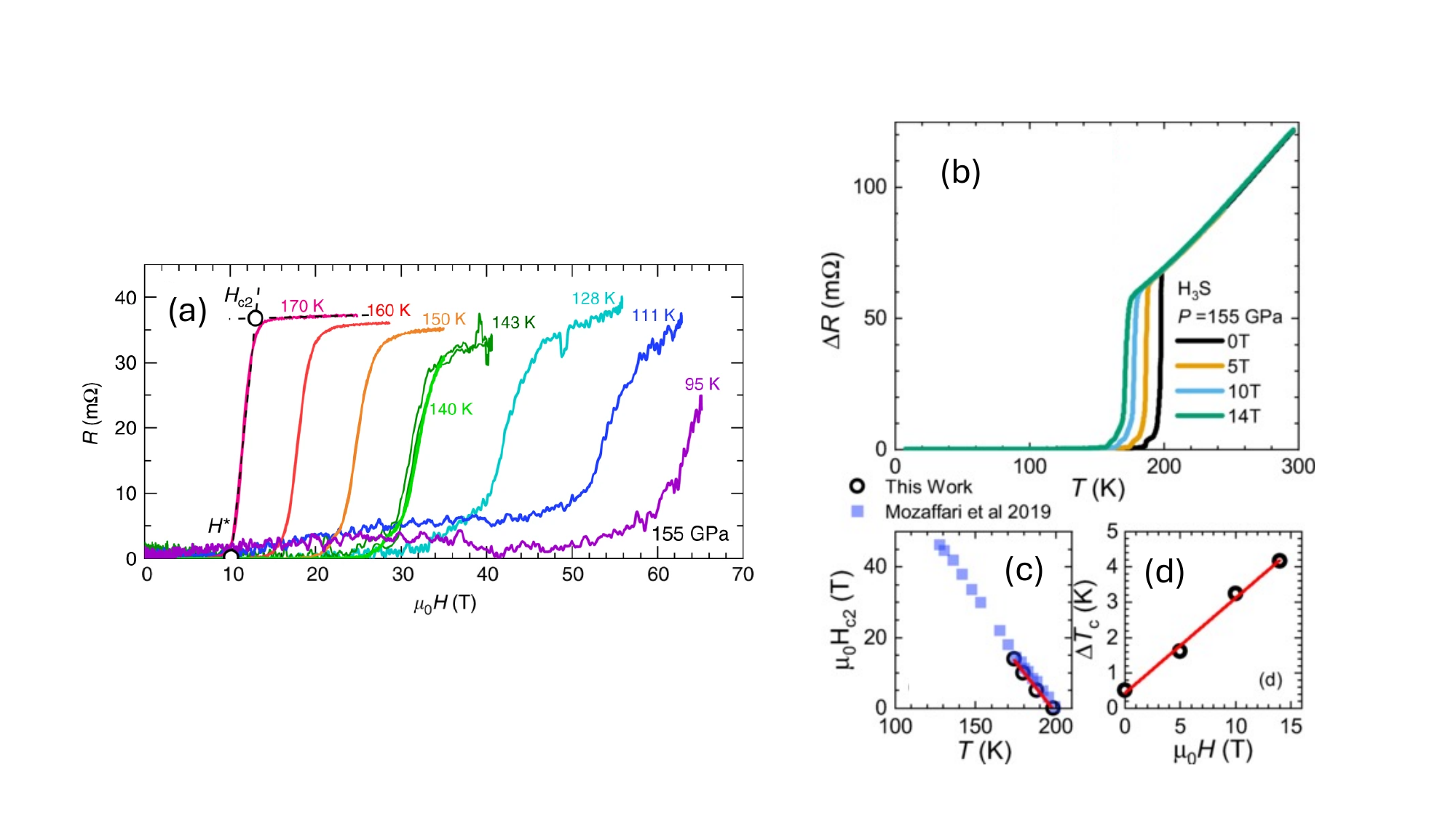}
\caption{ Resistive transitions for two samples of H$_3$S pressurized to $155\, {\rm GPa}$, a pressure at which transitions at approximately 200 K are seen.  In panel (a) the magnetic field dependence of the resistance is shown at a series of fixed temperatures studied to high fields in a DC magnet ($T > 145\, {\rm K}$) and pulsed-field magnet ($T < 145\, {\rm K}$); Ref.~\onlinecite{4}.  The higher noise levels in the pulsed field work are not unusual.  Panel (b) shows resistive transitions as a function of temperature for a series of fixed fields of 14 T and below\cite{5}, performed on a completely different sample from that in panel (a), as described in the main text.  A contribution attributed to sulphur surrounding the H$_3$S has been subtracted, so the data are shown as $\Delta R$ rather than as $R$. Panel (c) demonstrates excellent agreement of upper critical fields extracted from the two different experiments (shown in blue\cite{4} and open black circles\cite{5}), while panel (d) quantifies transition broadening under field of the experiment of Ref.~\onlinecite{5}. }
\label{fig1}
\end{figure}

For our current purpose of assessing experimental information, we restrict ourselves to two classic probes of superconductivity, resistance and magnetization, because superconducting responses to such probes are so well known.  This should not, however, be taken to imply any judgement on the work that is also going on to develop new measurement techniques specifically suited to the hydride sample environment\cite{2}.

The challenges of performing experiments on hydrides should not be underestimated.  Most of the groups involved have been very open about the nature of the high-pressure matter that is produced in such experiments.  It is chemically inhomogeneous, and the phases that exist in the sample are often hard to identify with certainty.  This chemical uncertainty is not surprising: it is an environment in which it is very difficult to use the standard synthesis and characterization techniques of solid-state chemistry.  In this sense the materials uncertainties associated with any new family of superconductors are higher in the hydrides than in any previously studied materials class.

For the purposes of analyzing data from physical measurements, the at-present unavoidable inhomogeneity must be borne in mind.  One inevitably expects variations in details of electrical resistance data, for example.  If there is superconductivity in such an environment, some transitions will be the result of establishing fragile percolation paths, while others will be incomplete because only a non-connected fraction of the sample is superconducting.  In the case of magnetic measurements, the sample environment gives a further series of challenges.  Even using the smallest cells specially built for the purpose, the mass of the cell is approximately $100$ million times larger than that of the potential superconductor inside, so extreme care must be taken to reduce background signals to the level where any superconducting contribution can be seen in the data.  

\section*{Measurements of resistance, upper critical field and magnetization}

We begin with an examination of the key resistive evidence for superconductivity.  Many measurements have been carried out using four-terminal measurements in which, for a homogeneous superconductor, the resistance would fall to the noise floor of the measurement apparatus in the superconducting state.  In some cases, for example Refs.~\onlinecite{1,2,3} this is seen, likely because a percolation path has been established through the pressurized material rather than because a homogenous sample has been realized.  In others, for example Refs.~\onlinecite{4,5}, zero resistance is not achieved, but there is evidence from real-space imaging for why a complete percolation path is unlikely.

In isolation, this kind of resistive evidence would not be sufficient for an assertion of superconductivity.  However, it is supplemented in the literature by numerous reports of the suppression of the resistive transition by an applied magnetic field. We reproduce two examples, from Refs.~\onlinecite{4} and ~\onlinecite{5}, in Fig.~\ref{fig1}.  

Several further aspects of the data shown in Fig.~\ref{fig1} merit comment.  Firstly, the data in Fig.~\ref{fig1}(a) were taken at the US National High Magnetic Field Laboratory in Tallahassee and Los Alamos, by a team led by scientists otherwise unconnected with the Mainz group in which the pressurized sample was prepared. Secondly, the sample studied in the work shown in Fig.~\ref{fig1}(b) was prepared by an entirely different group at the University of Bristol, using an entirely different synthesis route.  The critical temperatures reported for H$_3$S in the Mainz experiment\cite{1}, the Los Alamos experiment\cite{4} and the Bristol experiment\cite{5} agree within experimental error, as do the deduced upper critical fields of the Tallahassee/Los Alamos and Bristol studies (see Fig.~\ref{fig1}(c)).  This level of reproducibility of findings between different groups is one of the requirements for claims of any new phenomenon to be regarded as credible.  In this context we further note that the sample preparation and measurement routes used in Refs.~\onlinecite{1,2,3} were also independent.

\begin{figure}[ht]
\centering
\includegraphics[scale=0.40]{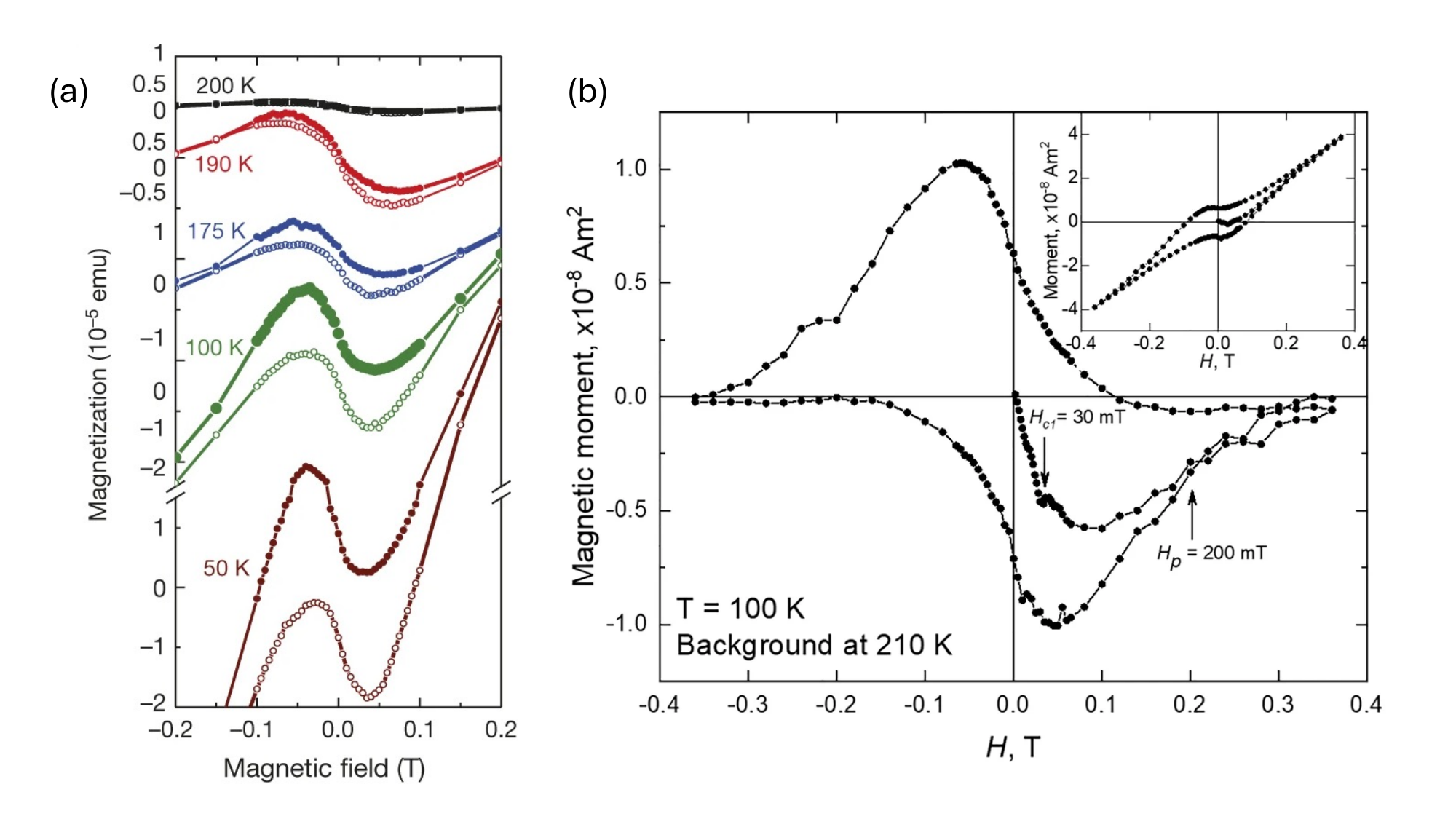}
\caption{ Magnetisation loops for two samples of H$_3$S pressured to $155\, {\rm GPa}$\cite{1} (a) and $140\, {\rm GPa}$\cite{6} (b) respectively, with the virgin curve included in (b).  The main feature of the loops are visible without any subtraction [(a) and inset to (b)].  In (a) the loop is also seen to close at $T_{\rm c}$ (black data), in line with expectations for a superconductor.  }
\label{fig2}
\end{figure}

Next, we turn our attention to SQUID measurements of magnetization.  As mentioned above, these are extremely challenging in an environment where the sample mass is necessarily one part in $10^8$ of the mass of the pressure cell, and have only been reported so far by the Mainz group.  The cell components, and minute levels of impurities in or on their surface, can give background signals (diamagnetic or paramagnetic, depending on details of the impurities and cell components).  In the original 2015 paper\cite{1}, magnetization loops were shown without the so-called ‘virgin curve’, which is the data seen on the first cycle of the loop following zero-field cooling.  Since then, data have been reported including virgin curves\cite{6}.  We reproduce some of those data  in Fig.~\ref{fig2}.  Panel (a) shows the data from Ref.~\onlinecite{1}, without background subtraction.  The hysteretic part of the signal disappears (i.e. the hysteresis loops close) at $T_{\rm c}$, fully consistent with their origin being superconductivity.  Panel (b) shows the loop reported in Ref.~\onlinecite{6}. Data including the background are shown in the inset.  Even before background subtraction, the data for the virgin curve go negative, in low applied fields, offering strong evidence that diamagnetism is observed in the raw data.  The simple subtraction of the linear background gives the curve shown in the main plot, which has the main qualitative features expected of a superconducting hysteresis loop in the presence of flux trapping.

\section*{Conclusion}

The goal of this brief article has neither been to review the whole field of hydride superconductivity nor to discuss the issues of detail about H$_3$S work raised in recent correspondence on ArXiv and other preprint servers and in the popular press.  It has been to assess the broader scientific question of whether hydride superconductivity is genuine or not. Based on the data we have shown and discussed here, in our professional judgement it is overwhelmingly probable that the superconductivity in this class of compounds is genuine.  It is also exciting and ground-breaking, making it even more important that data be made publicly available and subjected to reasonable scientific skepticism.  The most useful form for such skepticism to take is experiments attempting to confirm or deny those already performed, combined with others building on the existing knowledge and driving it forward by introducing new compounds and measurement techniques. We hope that by firmly endorsing the overall validity of the field, we can help promote continued and, indeed, enhanced experimental activity in it.  Our message to funding agencies is to continue to support good proposals to drive hydride superconductivity forward, and our message to young scientists is to enter the field with curiosity and enthusiasm if it is the kind of science that intrigues you. Finally, our message to the field’s pioneers is to congratulate and thank you for your important work.

\end{document}